# Social capital at venture capital firms and their financial performance: Evidence from China

Qi-lin Cao[a], Hua-yun Xiang[a], You-jia Mao[a], Ben-zhang Yang[b,1]

[a.]School of Business, Sichuan University, Chengdu, P.R. China, 610000

[b.]Department of Mathematics, Sichuan University, Chengdu, P.R. China, 610064

**Abstract**: This paper studies the extent to which social capital drives performance in the Chinese venture capital (VC) market and explores the trend toward VC syndication in China. First, we propose a hybrid model based on syndicated social networks and the latent-variable model, which describes the social capital at VC firms and builds relationships between social capital and performance at VC firms. Then, we build three hypotheses about the relationships and test the hypotheses using our proposed model. Some numerical simulations are given to support the test results. Finally, we show that the correlations between social capital and financial performance at VC firms are weak in China and find that China's VC firms lack mature social capital links.

**Keywords**: financial performance, latent-variable model, social capital, venture capital

In recent years, the number of venture capital (VC) firms that repeatedly turn to fundraising and investment to finance their projects by co-investing has grown rapidly (Aizenman and Kendall 2012; Guller and Guillén 2010; Meuleman and Wright 2011). Over the past two decades, many VC firms have leveraged their prominence to build co-investment networks (Dai et al. 2012; Jääskeläinen and Maula 2014). Using these networks, they aim to access higher investment returns and broader diversification opportunities in financial markets. Co-investment by different venture capitalists not only helps to establish the VC industry but also increases the rate of success of their projects (Rosiello et al., 2011).

---

[1] Corresponding author: yangbenzhang@126.com

In China, the VC industry has been developing for more than 20 years and especially rapidly in recent years. In particular, cooperation through co-investment, which is an investment strategy of VC institutions to diversify risk, share resources, and share value, has become an important method of investment in China to enhance investment performance. By retaining a significant degree of ownership after the firms in which they invest go public, VC firms can continue their monitoring role, which further accelerates the growth and performance of those recipient firms. The literature, however, shows that VC activities are conducted in a markedly different way in China from elsewhere. Bruton and Ahlstrom (2003) argue that this is because the institutional environment in China is different from that in the West. For example, weak enforcement of investor protections and poor corporate governance in China might reduce VC firms' incentives to act in the interests of investors (Bruton and Ahlstrom 2003; Lu et al. 2013). The institutional differences in China can lead to underpricing and different long-run performance outcomes. As in the literature, we define the social capital of each VC firm as everything that it leverages in its own resources to build co-investment networks. Also, the literature shows that different characteristics at VC firms affect the recipient firms' performance and that startup firms choose specific VC firms in order to signal their quality and reduce the information asymmetry between those recipient firms and other market participants (Cumming et al. 2006; Hsu 2004). But the development of co-investment has been relatively slow in recent years, which attracted our attention. The question of whether China's unique legal system and cultural background of co-investment can help to improve performance is one worth studying.

The motivation for our study is to test whether social capital at VC firms affects their investment performance. First, this study employs the social network method to describe the social capital of venture capital. Then, using a two-node network built by VSs and their co-investment projects, we discuss the contribution of a network of the projecting to VCs' part in a two-node network. We use a latent-variable model to test the effects of the social capital of venture capital on the success of venture capital in China. Unlike in classical research modeling performance and networks (Hochberg et al. 2007, 2010), we employ a latent-variable model with social capital: we did not consider what social capital includes, just what it represents in a network. This makes sense because the topic of this paper is the characteristics of China's VC, not the social capital of VC in general.

First, this study contributes to the literature by enhancing understanding of the corporate relationship between financial performance and the social capital of venture capital in China. Using social network attributes and other information, we test the latent-variable model to test our hypothesis. Second, the paper contributes new evidence regarding the investment phenomenon of venture capital in China. Finally, we offer some analyses and suggestions regarding venture capital in China.

The remainder of the paper is organized as follows. First, we give some background on venture capital and financial reporting in China. Then we review related literature, develop our hypotheses, and test our model. Sample data and descriptive statistics are presented in the following section, and then the statistical results are discussed. Finally, we present our conclusions.

## Venture Capital and Financial Reporting in China

According to reports from the Zero2IPO, 597 VC funds were added in China, with

30,700,000 dollars in new capital, 12in 2015 . New investment was made in 3,554 ventures and the scale of investment reached about 20 billion dollars in 2015. Investment is booming thanks to a policy of fiscal stimulus and extensive government spending, and the Chinese government recognizes VC investment as a vital way to support small companies and support economic growth. Indeed, unlike in mature markets, in which stimulus packages are usually market oriented, with well-designed mechanisms to assist in their implementation, in emerging markets, stimulus packages are often more intended to benefit SOEs and other state-controlled entities, rather than benefiting private firms. This is especially the case in China, where the government still controls a large number of firms and the banking system, which raises the question of whether Chinese state-owned enterprises (SOEs) and non-state-owned enterprises (non-SOEs) have equal access to the increased funding for bank loans and investment opportunities. For example, Leary (2009) indicates that, following an expansion (contraction) in the supply of bank loans, the ratio of bank loans to small, bank-dependent firms significantly increases (decreases) relative to that of large, less bank-dependent firms, because small, bank-dependent firms are less likely to switch between public and private debt markets. Lemmon and Roberts (2010) show that net debt issuances decline significantly after a contraction in the supply of bank credit. Shen et al. (2015) find that the economic stimulus package in China leads to better access to bank loans for large firms or SOEs than for small firms or privately owned firms. The investment made by Chinese firms come to a new development stage, after the implementation of the stimulus package, and private capital has become the main source of venture capital.

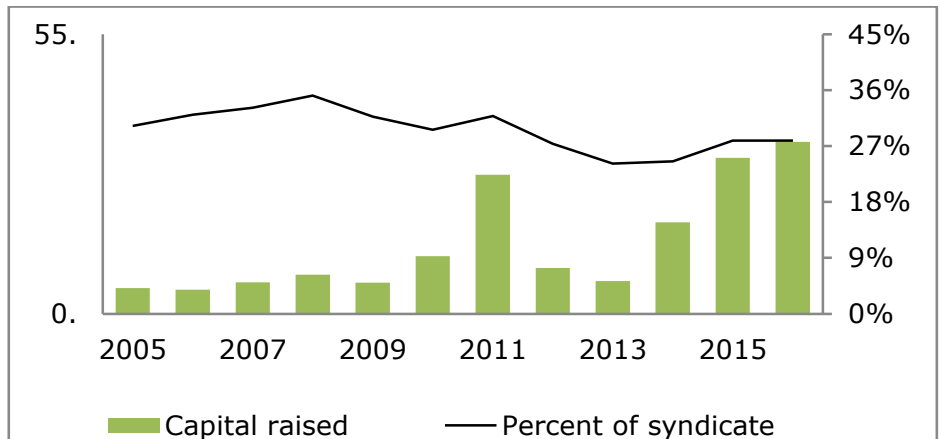

Figure 1. The incremental capital raised by VC institutions in China.

Note: The left-hand y-axis is the value of incremental capital raised, in billions of dollars; the right-hand y-axis is the percentage of VC syndication, by percentage of projects.

Figure 1 shows that in general the percentage of projects with VC syndication gradually decreased after the 2008 financial crisis. Despite the trend toward globalization, business cycles in industrialized countries and emerging Asian economies have so far remained largely independent of each other. This is referred to in the recent literature as a decoupling of business cycles (Akin and Kose 2008). Nevertheless, recent developments since the onset of the global financial crisis in the second half of 2008 also show that these countries are not autonomous. The International Monetary Fund (Akin and Kose 2008) argues that the current slowdown in the world economy could have a significantly larger impact on Asian economies than earlier global downturns because of their more extensive trade and financial integration, especially with the US. Furthermore, Hong et al. (2009) show in their historical analysis that earlier worldwide financial crises often had overwhelming impacts on Asian economies. Thus, the decline in the percentage of syndication may also be affected by the financial crisis. For risk control purposes, VC firms have become more careful and take a longer time to exit, thus their investment decisions have changed. Some small reverses appear in the years when a large amount of capital was raised (for political reasons). These reverses may have been caused by the excessive capital invested in high-risk projects, so VC firms syndicate to lower their risk. VC in joint investments is a hot issue and has many advantages, but the decline in VC syndication in China is inconsistent with the theoretical conclusion. Co-investment in China does not seem to be guided by rules about how syndicate members are chosen, and VC firms selecting a VC partner have a wide range of technologies, a wide range of information, industry expertise, and investment networks. Is VC syndication useful in China? Why is the percentage of syndicated projects declining?

**Prior Research and Development of Hypotheses**

Social capital describes the ability of individuals or groups to access resources embedded in their social network. Social capital can enrich other forms of capital, such as favors (human capital) or new information (intellectual capital). Lin et al. (2001) defines social capital as "investment in social relations with expected returns in the marketplace." In this framework, social capital is the creation of social interactions and the expectation of future social resources. The function of social capital is split into "bridging" and "bonding," which describe resources embedded in different types of relationships. The ties that connect different clusters within a network, often referred to as "bridging" ties, help propagate novel information across those groups. As explained below, bridging ties are more likely to be weaker ties and thus provide access to novel information and diverse perspectives. Stronger ties are multiple iterative interactions and higher levels of trust, support, and intimacy. These ties typically provide access to more substantive forms of capital conversion associated with bonded social capital. Facebook networks contain both strong and weak ties (Bakshy et al. 2012). Because individuals often use multiple channels to communicate with strong ties and fewer channels to interact with weaker ties, the focus of this study is on bridging social capital because we are interested in how Facebook enables greater access to resources held by weaker ties, which may not be available through other channels.

VC firms have a social capital network, which is of great help to enterprises. Indigenous VC energizes the construction of social relations networks (Pruthi et al. 2003), which play a role similar to that of VC as the representative of the financing side and enterprise ecosphere. Basalp (2011) constructs a network model of the science and technology park; they finds in the establishment of a cooperative, innovation and information communication network in the process of playing an intermediary role. The number of partners does not affect the corresponding social capital. Recent research on social capital extends to the meso-structure of relationship networks among firms. Block et al. (2012), using data analysis find that the social capital structure and relationship of VC firms have significantly effects on VC financing for startups. The exchange of social capital within the network plays a significant role in shaping the network's structure (Jackson et al. 2012). From the perspective of enterprises, the allocation of social capital and network structure directly enables enterprises to obtain financial resources, which reflects enterprises' information dissemination in the VC network (Chakir 2013). Alexy et al. (2012) believe that the social capital possessed by venture capitalists allows them to obtain superior access to information about investment. According to Javakhadze et al. (2016), social capital will reduce the negative impact of incomplete information and agency problems and affect external finance and cash flow as well as external finance and Tobin'Q sensitivities

A VC co-investment network is formed out of a pattern of social capital exchange between VC firms and projects. Wilson (1968) define an investment alliance as "a group of independent decision makers who make common decisions in an uncertain situation and share a common benefit" ( p.119) . The motivation and effect of joint investment are the two main areas of VC cooperation. Joint investment has various motivations. The value-added hypothesis holds that different VC firms, with different management skills, can increase the value of the investment. The joint investment choice hypothesis posits that the co-investment gains advantage if many uncorrelated VC firms evaluate the same project (Casamatta and Haritchabalet 2007). There are also many other hypotheses, such as those on decentralized financial risk, information sharing and potential investment opportunities, social structural factors, and window dressing (Lockeett and Wright 2001). Previous research proves the relationship between social capital and social networks. But joint investment networks consists of firms and projects need to be transformed into a VC firm-VC firm form, so that a social network analysis method can be used to construct the network consist of like elements (firm to firm), not different ones (firm to project). Evidence on VC networks shows that the more central the location of a network, the greater the advantage in interaction and the higher the VC firm's performance (Checkley 2014; Tastan 2013).

## Hypotheses and methodology

Most studies employ non-network methods to analyze social capital. A VC syndication network can be used to measure social capital. The structural hole of VC firms represents the power and opportunity of social capital. The closer the network between the VC firms is, the more obvious the network characteristics are, and the more obvious the network effect of cooperation is, the better the embedded network. Resources can flow smoothly (through shorter paths) between VC firms. For individual venture capitalists, the resources and advantages of other venture capitalists comprise social capital.

*Hypothesis 1: The higher the network embeddedness index and the structural hole index of VC institutions, the higher the social capital of VC firms.*

Outside China, most VC firms are independent, they pursue financial returns, and their success can be measured by their financial performance. However, for corporate VC firms, financial returns are not the only or even the primary goal; rather, they may be seeking the completion of an initial public offering or an acquisition at an attractive price, so an initial public offering (IPO) is a measure of failure, rather than success (Gompers and Lerner 2000). Due to the emerging firms' high Price-Earnings Ratio, an IPO is much more profitable than a merger or a buyout in China. Even for corporate VC firms, going public is a kind of success, as long as projects are operated independently. IPO exits of VC firms often achieve higher exit returns than other exit ways such as merger and acquisition, management buy-outs. It is widely accepted that the higher a firm's proportion of IPO exits, the higher the firm's returns on its VC exit. When VC has a higher exit ratio, the financial gains will be higher. This means VC is closer to obtain a better performance.

*Hypothesis 2: The higher the financial returns, the higher the performance of the VC firm.*

Many researchers have shown that social capital through social networks creates channels for the flow of information and plays an important role in portfolio and market participation (Cohen et al. 2008; Fafchamps, 2007; Hochberg et al. 2007). Therefore, social capital has an important impact on corporate finance, especially external financing, while trust can effectively operate in economic entities and reduce transaction costs (Javakhadze et al. 2016). Social capital can also increase economic efficiency and encourage economic entities to build reputations for good-faith trading through a reputation-loss disciplinary mechanism (McMillan and Woodruff 2000).

If the VC network really has meaning, as in the previous literature, network embeddedness is helpful for VC investment and value-added services. Thus the project is more likely to succeed. As the resources obtained through the network can be defined as social capital, the more abundant the social capital is, the more successful the VC investment project will be. The higher the project success ratio and the more successful the project is, the more successful the VC institution will be. The idea of social capital supports VC success, not project success; VC success may not be the only measure of project success but, rather, more projects, more stable returns, and so forth. The motivation for our study is to test whether the social capital of VC firms affects the success of projects in which they invest. Therefore, we propose the following hypothesis:

*Hypothesis 3: The greater the social capital of VC firms, the better their performance is.*

Based on this logic, we can study how social capital affects VC success using the latent-variable model. First, as Figure 2 shows, the relationship between VC firms and projects is described using a two-node social network (bipartite network). Bipartite networks have two node types, such that two nodes can be linked only if they are of different types. More precisely, the network has two dimensions: VC and the project. As the special and determined choices of each VC firm, VC and projects are linked by different connections.

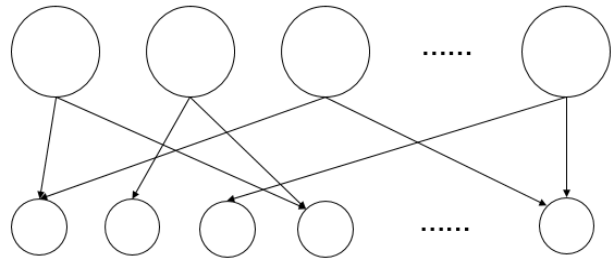

Figure 2. Sketch of joint investment two-node social network.

Second, studying the VC social capital network relies on a network model of the second node. This model consists of two classes of nodes: participants and events. In this study, the participants in the network are VC institutions, and the events are venture capitalists. Generally, to obtain the one part's attribute, mapping to a one-node network for analyze is a useful method. As shown in Figure 3, projecting to the VC part, we obtain the VC syndication network.

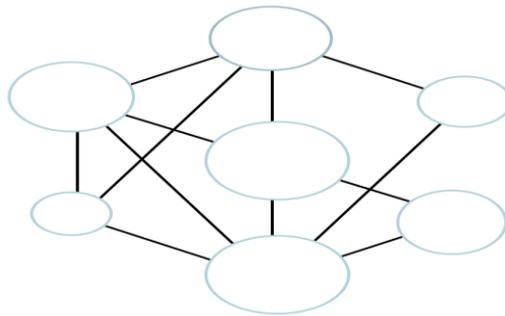

Figure 3. Structure of the VC network

Then we use a social network analysis (SNA) to obtain social capital variables and calculate VC firms' financial performance. Centrality, as measured in SNA, indicates how network is connected, and constraint indicate the structural holes in network that are all traditional social capital measures. Degree is a local centrality measure, while closeness and betweenness are global centrality measures. Some social capital, like physical resources, will not exceed social ties, but information will come much farther, and the network location is time related. They are node-level characteristics, a VC's direct benefit (embeddedness) and indirect benefit (structural holes) from network embeddedness. Scale and return are included in financial performance. Scale is the result of VC financing and may influence future financing.

Finally, we use the latent-space approach to model social capital and VC performance over time. We investigate how social capital affects performance in the following latent-variable model:

$$\text{Social Capital} = \alpha_1 \times \text{Weighted degree} + \alpha_2 \times \text{Closeness} + \alpha_3 \times \text{Betweenness} + \alpha_4 \times \text{Structural hole} + \varepsilon_1,$$

$$\text{Performance} = \beta_1 \times \text{Investment in} + \beta_2 \times \text{IPO proportion} + \beta_3 \times \text{Weighted book return} + \beta_4 \times \text{Weighted IRR} + \beta_5 \times \text{Investment exited} + \beta_6 \times \text{Exit rate} + \varepsilon_2,$$

and structural equation

$$\text{Performance} = \eta \cdot \text{Performance} + \gamma \cdot \text{Social capital} + \varepsilon_3,,$$

where $\alpha_i, \beta_i, \eta, \gamma$ are coefficients and $\varepsilon_i$ are error terms.

Indeed, latent-variable models go beyond ordinary regression models to incorporate multiple independent and dependent variables as well as hypothetical latent constructs that clusters of observed variables might represent. This approach uses the idea that to test the specified set of relationships among observed and latent variables as a whole and allows theoretical testing even when experiments are not possible. For social capital, the latent variables of social network features are described for VC, which means that the social capital of VC is in their social networks. In this paper, we use the degree of weight, closeness, betweenness, and structural holes as observed variables of VC. An additional important aspect of our model is VC performance. We similarly use the observed variables of financial returns and only financial returns as permanent VC. Separating the relationship between latent variables, we build an equation of two latent variables: social capital and performance. In addition, we do not include industry concentration, professionalization, VC lifetime, etc., which are part of social capital and represented in social networks. In this model, we are not concerned about what social capital is, only what it represents in networks. The performance is pure financial performance, not including accumulation of social capital.

To test our three hypotheses, we construct four latent-variable models, in which the model described above is Model 1, and Models 2, 3, and 4 control for a structural hole variable, variables for both inward investment and outward investment, and the three variables mentioned above, respectively.

## Data

The data are collected from Qingke and Wind, which are leading financial information providers in China. The sample period is from 2000 to 2015. First, as Table 2 shows, we obtained the data from Zdatabase disclosed from 2000 to 2015 for 4,985 VC firms, 21,421 enterprises with VC investment, and 40,882 investment events in China, omitting undisclosed data.

Certain investments may be more or less uncertain to different VC firms, based on the extent of their prior experience in the specific industry of the investment. We thus focus on a hitherto unexplored, relational aspect of uncertainty that is inherent in a particular VC firm–investment dyad and reflects the VC firm's understanding of the particular company and its environment. In view of this, the need and opportunity to syndicate a particular investment may not always be aligned: in certain situations— such as when the investment is novel for the VC firm—the factors that increase the firm's need to seek partners may also reduce the firm's opportunity to find and attract needed partners. Such a misalignment is related to the presence of two different types of uncertainty. On the one hand, a VC firm faces (egocentric) uncertainty related to the proper decisions to be made in selecting and managing these investments. In these cases, the firm can benefit from the participation of syndicate partners and is thus likely to seek such partners. Yet, on the other hand, for these very same investments, it is difficult for potential syndicate partners to evaluate the focal VC firm as a worthy partner (altercentric uncertainty). In these cases, the opportunity for syndication is contingent upon the VC firm's ability to signal its quality to potential partners, such as through its status and reputation. Thus, we divided the experiment into two situations: same-round investment and different-round investment, so we can separate the effects of the uncertainty caused by investment and compare performance under weak conditions.

We model social networks, in which raw data is modeled into a "two-node" network and then transformed into a "one-node" network. Thus, by modeling joint investment events into two-node networks, we construct a "same-round model" by considering joint investment as two VC firms sharing a common finance round, which means that VC firms meet and discuss terms directly with the companies they are funding. In the same-round model, VC firms have a strong relationship in the network, and their prices, costs, and benefits are all the same. We also construct a "different-round model" by considering joint investment as two VC investments by the same company, which mean VC firms are all shareholders or even board members of the same company, so their interests are same but the relationships are weaker than in the "same-round model" and contain both "strong ties" and "weak ties."

Next, with Pajek 4.10 software, we processed the investment relationship into the pattern of a two-node network and then mapped it into a one-node network. Removing the isolates, we created a VC syndication network composed of 3,694 VC firms. After constructing syndication networks, we selected degree centrality, closeness centrality, betweenness centrality, and structural holes as network indicators to outline the social capital of VC firms. We obtained 1,115 exit events from 2000 to 2015 and defined six indicators to characterize VC firms' investment performance. Table 1 contains all variable definitions, Table 2 shows the General situation of the sample, and Table 3 summarizes and gives numerical descriptions of these variables.

Table 1. Definition of variables in the model with expected sign

| Variable name | Definition | Expected sign |
|---|---|---|
| Weighted degree | Sum of degrees of each node in co-investment network | |
| Closeness | Sum of the shortest distances from a node to another node in the network | |
| Betweenness | Degree of resource control of a node in a network. | |
| Structural hole | Gaps in the social network, measured by constraints | |
| Investment (total) | Sum of investments by each VC (including those are not withdrawn) | Positive |
| IPO proportion | Ratio of IPO amounts to total exits. | Positive |
| Weighted book return | Book return weighted by total investment of each project | |
| Weighted IRR | Internal rate of return weighted by total investment of each project | |
| Investment exited | Sum of investment that finally exits | Positive |
| Exit ratio | Ratio of total investment to total investment entered | Positive |

Table 2. Sample selection procedure

| | |
|---|---|
| Sample period | 2000 to 2015 |
| VC firms | 4,985 |
| Enterprises with VC investment | 21,421 |
| Investment events | 40,882 |
| VC firms in joint investment networks | 3,964 |
| Disclosed exit events | 1,115 |

Table 3. Descriptive statistics of the variables

| | Mean | Median | Std. dev. | Min | Max | Q1 | Q3 |
|---|---|---|---|---|---|---|---|
| Weighted degree | 0.011 | 0.004 | 0.047 | 0.000 | 0.099 | 0.003 | 0.011 |
| Closeness | 0.005 | 0.002 | 0.032 | 0.001 | 1.000 | 0.000 | 0.003 |
| Betweenness | 0.005 | 0.002 | 0.032 | 0.000 | 1.000 | 0.000 | 0.003 |
| Structural hole | 0.381 | 0.340 | 0.249 | 0.000 | 0.900 | 0.163 | 0.720 |
| Investment (total) | 0.024 | 0.020 | 0.073 | 0.000 | 1.000 | 0.001 | 0.043 |
| IPO proportion | 0.848 | 0.700 | 0.250 | 0.003 | 1.000 | 0.694 | 1.000 |
| Weighted book return | 0.016 | 0.007 | 0.045 | 0.000 | 1.000 | 0.004 | 0.014 |
| Weighted IRR | 0.003 | 0.002 | 0.030 | 0.000 | 1.000 | 0.002 | 0.002 |
| Investment exited | 0.035 | 0.026 | 0.042 | 0.025 | 1.025 | 0.025 | 0.029 |
| Exit ratio | 0.001 | 0.001 | 0.000 | 0.000 | 1.000 | 0.000 | 0.001 |

*Note:* For definitions of the variables, see Table 1.

## Empirical Results

Table 4 summarizes the statistics of latent variables in same-round investment. If we compare the results of models 1 and 2, we can see that, in the same investment, social capital is deeply dependent on the weighted degree and structural hole. Therefore, the latent-variable model recognizes that when the network embeddedness index and the structural hole index of VC institutions increases, the social capital of venture capital will rise as well. Thus H1 can be tested immediately. Models 1 and 3 in Table 4 further indicate that the higher the financial returns, the higher the performance of the VC. This confirms H2. In addition, model 4 shows that the relationships heavily depend on three variables: structure hole, investment (total), and investment exited. This means that structural holes of VC networks can affect social capital, and VC performance is determined by investment (total) and investment exited variables.

Table 4. Results of latent variables in same-round investment

|  | Model 1 | Model 2 | Model 3 | Model 4 |
|---|---|---|---|---|
| *Social capital* | | | | |
| Weighted degree | 1.000 | 1.000 | 1.000 | 1.000 |
| Closeness | 0.567*** | 0.370*** | 0.572*** | 0.370*** |
|  | (25.089) | (26.178) | (24.781) | (26.095) |
| Betweenness | 0.568*** | 0.371*** | 0.572*** | 0.371*** |
|  | (25.110) | (26.321) | (24.800) | (26.237) |
| Structural hole | 0.578*** |  | 0.583*** |  |
|  | (25.282) |  | (24.950) |  |
| *Performance* | | | | |
| Investment (total) | 1.000*** | 1.000*** |  |  |
|  | (6.639) | (6.675) |  |  |
| IPO proportion | 0.450*** | 0.447*** | 1.001*** | 1.001*** |
|  | (9.298) | (9.347) | (47.236) | (46.762) |
| Weighted book return | 0.522*** | 0.522*** | 1.163*** | 1.172*** |
|  | (9.024) | (9.082) | (29.489) | (29.936) |
| Weighted IRR | 0.428*** | 0.424*** | 0.952*** | 0.950*** |
|  | (9.282) | (9.327) | (45.574) | (44.780) |
| Investment exited | 0.534*** | 0.531*** |  |  |
|  | (9.233) | (9.284) |  |  |
| Exit ratio | 0.476*** | 0.473*** | 1.059*** | 1.061*** |
|  | (9.262) | (9.313) | (43.232) | (43.226) |

*, ** and *** denote significance levels of 10%, 5% and 1% respectively.

The performance variable is significantly and positively associated with investment variables in the models. These results provide strong evidence that the money invested has the most impact on performance in China. Previous studies of Chinese investment performance did not recognize the importance of the role of the money invested and the weighed book return in their analysis of performance determinants, and thus the results of their comparison of investment performance in China may be misleading. With respect to other variables, as shown in Figure 5, the investment performance is consistent with the IPO proportion and weighted IRR in all three regression models. They all have a positive impact on performance. In same-round investment, the average coefficient of the weighted book return with performance is 0.844. This result is consistent with China's investment environment, because not all successful exit projects are good investments.

Table 5. Results of covariances and variances in same-round investment

| | Model 1 | Model 2 | Model 3 | Model 4 |
|---|---|---|---|---|
| Social capital performance | 0.032 | 0.034 | 0.014 | 0.015 |
| Variance | | | | |
| Weighted degree | 0.018*** (16.314) | 0.003*** (10.826) | 0.018*** (16.296) | 0.003*** (10.785) |
| Closeness | 0.001*** (14.892) | 0.004*** (16.151) | 0.001*** (14.746) | 0.004*** (16.165) |
| Betweenness | 0.001*** (14.870) | 0.004*** (16.145) | 0.001*** (14.725) | 0.004*** (16.159) |
| Structural hole | 0.001*** (14.673) | | 0.001*** (14.533) | |
| Investment (total) | 0.259*** (16.554) | 0.258*** (16.543) | | |
| IPO proportion | 0.001*** (15.064) | 0.001*** (14.569) | 0.001*** (14.969) | 0.001*** (14.417) |
| Weighted book return | 0.006*** (16.225) | 0.006*** (16.082) | 0.006*** (16.198) | 0.006*** (16.032) |
| Weighted IRR | 0.001*** (15.290) | 0.001*** (14.924) | 0.001*** (15.181) | 0.001*** (14.754) |
| Investment exited | 0.003*** (15.712) | 0.002*** (15.395) | | |
| Exit ratio | 0.002*** (15.506) | 0.002*** (15.114) | 0.002*** (15.426) | 0.002*** (14.974) |
| Social capital | 0.024*** (10.569) | 0.038*** (15.178) | 0.024*** (10.466) | 0.038*** (15.179) |
| Performance | 0.042*** (4.517) | 0.042*** (4.540) | 0.008*** (14.846) | 0.008*** (14.819) |

*, ** and *** denote significance levels of 10%, 5% and 1% respectively.

Table 5 depicts the relationship between social capital and financial performance at VC firms with different models. The average relationship is about 0.024, which implies a weak relationship between social capital and financial performance. Thus, co-investment in China does not seem to be guided by rules when choosing syndicate members: VC institutions that choose a VC partner have a wide range of technologies, a wide range of information, industry expertise, and investment networks. VC syndication in China may not receive abundant benefits from their social networks. That is why the percentage of syndicate projects declining.

Table 5 also tells us that the degree in the same-round model has less factor loading, direct connections that are all strong ties, is less important than a mix of connections. This supports the theory of "informational benefit" in China. Total investment in the same-round model has more factor loading in VC firm performance. This means that a VC rich in social capital may benefit more in scale than in returns, or, on the contrary, scale rather than returns will lead VC firms to have more social capital.

In order to compare our results with those in different-round investment, we import the various samples to complete the experiments. As in the analysis above, the determinants of social capital are the weighted degree and structural holes in social networks. The variable affecting the joint investment relationship in China is deeply related to VC firms that have a strong ability to connect to other elements. Tables 6 and 7 show that investment is critical to VC performance under robust testing. Thus, we can confirm the correctness of H1 and H2.

Table 6. Results of latent variables in different-round investment

| | Model 1 | Model 2 | Model 3 | Model 4 |
|---|---|---|---|---|
| *Social capital* | | | | |
| Weighted degree | 1.000 | 1.000 | 1.000 | 1.000 |
| Closeness | 0.488*** | 0.486*** | 0.494*** | 0.493*** |
| | (16.449) | (16.530) | (16.179) | (16.222) |
| Betweenness | 0.485*** | 0.482*** | 0.491*** | 0.489*** |
| | (16.425) | (16.505) | (16.158) | (16.199) |
| Structural hole | 0.494*** | | 0.499*** | |
| | (16.473) | | (16.198) | |
| *Performance* | | | | |
| Investment(total) | 1.000*** | 1.000*** | | |
| | (6.333) | (6.345) | | |
| IPO proportion | 0.464*** | 0.463*** | 1.002*** | 1.002*** |
| | (8.867) | (8.884) | (44.324) | (44.123) |
| Weighted book return | 0.527*** | 0.526*** | 1.141*** | 1.143*** |
| | (8.655) | (8.673) | (29.780) | (29.854) |
| Weighted IRR | 0.444*** | 0.443*** | 0.960*** | 0.960*** |
| | (8.856) | (8.872) | (43.176) | (42.991) |
| Investment exited | 0.547*** | 0.546*** | | |
| | (8.824) | (8.840) | | |
| Exit ratio | 0.514*** | 0.513*** | 1.112*** | 1.112*** |
| | (8.814) | (8.832) | (38.909) | (38.880) |

*, ** and *** denote significance levels of 10%, 5% and 1% respectively.

Table 7. Results of covariances and variances in different-round investment

|  | Model 1 | Model 2 | Model 3 | Model 4 |
|---|---|---|---|---|
| Social capital performance Variance | 0.032 | 0.033 | 0.015 | 0.015 |
|  | Estimate | Std.Err | Z-value | P(>\|z\|) |
| Weighted degree | 0.056*** | 0.056*** | 0.057*** | 0.056*** |
|  | (16.449) | (16.428) | (16.446) | (16.421) |
| Closeness | 0.001*** | 0.001*** | 0.001*** | 0.001*** |
|  | (14.390) | (13.090) | (14.297) | (12.983) |
| Betweenness | 0.001*** | 0.001*** | 0.001*** | 0.001*** |
|  | (14.493) | (13.282) | (14.399) | (13.172) |
| Structural hole | 0.001*** |  | 0.001*** |  |
|  | (14.270) |  | (14.198) |  |
| Investment (total) | 0.238*** | 0.237*** |  |  |
|  | (16.552) | (16.549) |  |  |
| IPO proportion | 0.001*** | 0.001*** | 0.001*** | 0.001*** |
|  | (15.011) | (14.886) | (14.906) | (14.766) |
| Weighted book return | 0.005*** | 0.005*** | 0.005*** | 0.005*** |
|  | (16.140) | (16.097) | (16.102) | (16.051) |
| Weighted IRR | 0.001*** | 0.001*** | 0.001*** | 0.001*** |
|  | (15.195) | (15.087) | (15.073) | (14.944) |
| Investment exited | 0.002*** | 0.002*** |  |  |
|  | (15.541) | (15.452) |  |  |
| Exit ratio | 0.002*** | 0.002*** | 0.002*** | 0.002*** |
|  | (15.613) | (15.525) | (15.540) | (15.438) |
| Social capital | 0.030*** | 0.030*** | 0.029*** | 0.030*** |
|  | (7.591) | (7.622) | (7.482) | (7.499) |
| Performance | 0.035*** | 0.035*** | 0.008*** | 0.008*** |
|  | (4.323) | (4.331) | (14.647) | (14.633) |

*, ** and *** denote significance levels of 10%, 5% and 1% respectively.

Another question is whether the VC network really has meaning and whether network embeddedness will help VC investment and value-added services, making the project more likely to obtain better financial performance. As Tables 6 and 7 show, in different-round investment, VC firms' performance likely does not depend on their social capital, which may lead to the decline of VC syndication in China. One interesting phenomenon is that, in different-round investment, some variables may not have a major influence on the relationship between social capital and financial performance at VC firms. Indeed, for risk control purposes, VC firms have become more careful in different-round investment, and their investment decisions may change compared to those in same-round investment.

## Conclusion

Using a latent-variable model, we build three hypotheses about the relationships between social capital and VC performance in China. We find that the degree-centrality embeddedness in co-investment social network is much more important than other embeddedness variables in social capital that lead to VC success. Total investment (total scale of funds) is much more important than other performance variables. The explicit number of social relations and local embeddedness of a VC firm have a higher factor loading of real effective social relations. VC success brought about by social capital is mainly reflected in the correlation with VC's investment scale, rather than financial returns. Every variable indicates that, in China, social capital that leads to success is not about the actual performance of the best VC firms, but how good they appear to be. This may not be intentional but, rather, be due to information asymmetry.

Social capital has a weak relationship with VC firm success, although VC firm success owes more to fund size than to IPO or better returns. Chinese data on the relationship between social capital and VC firm success does not clearly support the logic that social capital contributes to the exit rate of VC IPOs or the rate of returns. It may be caused by other factors, such as economies of scale. It is also possible that VC firm success (on the surface) creates more social capital (with more connections). In conclusion, Chinese VC firms operate imperfectly. Although Chinese VC is one of the largest markets in the world, it remains young and immature. This may be why the percentage of VC syndicated is slowly decreasing in China, giving way to what really matters.

This paper does not directly consider specialization among VC institutions. The degree of concentration of VC institutions, that is, the extent to which a VC institution deals with a particular industry, may produce more specialized knowledge, which gives one VC firm an advantage over others. More importantly, the utility of expertise comes not from social capital but the mutual benefit of cooperation, which is embodied in network embeddedness. Therefore, it is not very meaningful to discuss the professionalism of individual VC firms.

## References

Aizenman, J., and Kendall, J. 2012. "The Internationalization of Venture Capital." *Journal of Economic Studies* 39 (5): 488–511.

Akin, C., and Kose, M. A. 2008. "Changing Nature of North-South Linkages: Stylized Facts and Explanations." *Journal of Asian Economics* 19 (1): 1–28.

Alexy, O.T., Block, J.H., Sandner, P., and TerWal, A. L. J. 2012. "Social Capital of Venture Capitalists and Start-Up Funding." *Small Business Economics* 39(4): 835–51.

Bakshy, E., Rosenn, I., Marlow, C., et al. 2012. "The Role of Social Networks in Information Diffusion." Proceedings of the 21st International Conference on

World Wide Web, *ACM*, 519-528

Basalp, S.G.G. 2011. "Establishing a Network among Entrepreneurs, Angle Investors and Venture Capital Firms: Structure and Importance of This Network." *Business and Economics Research Journal* 2: 153-64.

Block, J.H., et al. 2012. "Social Capital of Venture Capitalists and Start-Up Funding." *Small Business Economics* 39(4): 835-51.

Bruton, G.D., and Ahlstrom, D. 2003. "An Institutional View of China's Venture Capital Industry: Explaining the Difference between China and the West." *Journal of business venturing* 18 (2): 233–59.

Casamatta, C., and Haritchabalet, C. 2007. "Experience, screening and syndication in venture capital investments." *Journal of Financial Intermediation* 16(3): 368-398.

Chakir, B.B.A. 2013. "Entrepreneurs Access to Venture Capital in Moroccans Technology-Based Ventures: An Exploratory Study of the Role of Social Capital." *International Journal of Business and Social Science* 4(8):144-161

Checkley, M., Steglich, C., Angwin, D., and Endersby, R. 2014. "Firm performance and the evolution of cooperative interfirm networks: UK venture capital syndication." *Strategic Change* 23(1‐2): 107-118.

Cohen, L., Frazzini, A., and Malloy, C. 2008. "The small world of investing: Board connections and mutual fund returns." *Journal of Political Economy* 116(5): 951-979.

Cumming, D., Fleming, G., and Schwienbacher, A. 2006. "Legality and Venture Capital Exits." *Journal of Corporate Finance* 12(2): 214–245.

Dai, N., Jo, H., and Kassicieh, S. 2012. "Cross-Border Venture Capital Investments in Asia: Selection and Exit Performance." *Journal of Business Venturing* 27(6):

666–684.

Fafchamps, M., and Gubert, F. 2007. "Risk sharing and network formation." *American Economic Review* 97(2): 75-79.

Gompers, P., and Lerner, J. 2000. "The determinants of corporate venture capital success: Organizational structure, incentives, and complementarities." In Concentrated corporate ownership (pp. 17-54). University of Chicago Press.

Guller, I., and Guillén, M. 2010. "Home Country Networks and Foreign Expansion: Evidence from the Venture Capital Industry." *Academy of Management Journal* 53 (2): 390–410.

Hochberg, Y. V., Ljungqvist, A., and Lu, Y. 2007. "Whom You Know Matters: Venture Capital Networks and Investment Performance." *Journal of Finance* 62(1): 251–301.

———. 2010. "Networking as a Barrier to Entry and the Competitive Supply of Venture Capital." *Journal of Finance* 65(3): 829–859.

Hong, K., Lee, J.-W., and Tang, H. C. 2009. "Crises in Asia: Historical Perspectives and Implications." *Journal of Asian Economies* 21(3): 265–279.

Hsu, D. 2004. "Why Do Entrepreneurs Pay for Venture Capital Affiliation?" *The Journal of Finance* 59(4): 1805–1844.

Jääskeläinen, M., and Maula, M. 2014. "Do Networks of Financial Intermediaries Help Reduce Local Bias? Evidence from the Cross-Border Venture Capital Exits." *Journal of Business Venturing* 29 (5): 704–721.

Jackson, M.O., Rodriguez-Barraquer, T., and Tan, X. 2012. "Social Capital and Social Quilts: Network Patterns of Favor Exchange." *American Economic Review* 102(5): 1857-1897.

Javakhadze, D., Ferris, S.P., and French, D.W. 2016. "Social Capital, Investments, and

External Financing." *Journal of Corporate Finance* 37: 38–55.

Leary, M. 2009. "Bank Loan Supply, Lender Choice, and Corporate Capital Structure." *Journal of Finance* 64 (3): 1143–85.

Lemmon, M., and Roberts, M.R. 2010. "The Response of Corporate Financing and Investment to Changes in the Supply of Credit." *Journal of Financial Quantitative Analysis* 45 (3): 555–87.

Lin, N. 2001. "Social Capital: Theory and Research." *Contemporary Sociology* 2: 1-38

Lockett, A., and Wright, M. 2001. "The syndication of venture capital investments." *Omega* 29(5): 375-390.

Lu, H., Tan, Y., and Huang, H. 2013. "Why Do Venture Capital Firms Exist: An Institution-Based Rent-Seeking Perspective and Chinese Evidence." *Asia Pacific Journal of Management* 30(3): 921–936.

McMillan, J., and Woodruff, C. 2000. "Private order under dysfunctional public order." *Michigan law review* 98(8): 2421-2458.

Meuleman, M., and Wright, M. 2011. "Cross-Border Private Equity Syndication: Institutional Context and Learning." *Journal of Business Venturing* 26(1): 35–48.

Pruthi, S., Wright, M., and Lockett, A. 2003. "Do Foreign and Domestic Venture Capital Firms Differ in Their Monitoring of Investees?" *Asia Pacific Journal of Management* 20(2): 175–204.

Rosiello, A., Avnimelech, G., and Teubal, M. 2011. "Towards a Systemic and Evolutionary Framework for Venture Capital Policy." *Journal of Evolutionary Economics* 21 (1): 167–189.

Shen, J., Firth, M., and Poon, W. 2015. "Bank Loan Supply and Corporate Capital Structure: Recent Evidence from China." Working paper, Hang Seng

Management College, Hong Kong, China.

Wilson, R. 1968. "The theory of syndicates." *Econometrica: journal of the Econometric Society* 36(1): 119-132.

Tastan, M., Falconieri, S., and Filatotchev, I. 2013. "Does venture capital syndicate size matter?." In European Financial Management Association 2013 Annual Meeting.